\documentclass [12pt, one column, a4paper] {article}

\title {Radiative Corrections to Casimir Effect in the $\lambda\phi^{4}$ Model}

\author{R.\ M.\ Cavalcanti\footnote{e-mail: \tt rmoritz@if.ufrj.br}, C.\ Farina\footnote{e-mail: \tt farina@if.ufrj.br}\\
{\small Instituto de F\'\i sica - Universidade Federal do Rio de Janeiro}\\
{\small Caixa Postal 68528 - CEP 21945-970, Rio de Janeiro, Brazil.}\\
and F. A. Barone\footnote{e-mail: \tt fbarone@feg.unesp.br}\\
{\small UNESP-Campus de Guaratinguet\'a, Departamento de F\'\i sica e Qu\'\i mica, Av. Ariberto Pereira}\\
{\small da Cunha 333, 12500-000, Guaratinguet\'a, SP, Brazil.}}

\date {}

\begin {document}

\baselineskip=20pt

\maketitle

\begin{abstract}
	We calculate radiative corrections to the Casimir effect for the massive complex scalar field with the $\lambda\phi^{4}$ self-interaction in $d+1$ dimensions. We consider the field submitted to four types of boundary conditions on two parallel planes, namely: (i) quasi-periodic boundary conditions, which interpolates continuously periodic and anti-periodic ones, (ii) Dirichlet conditions on both planes, (iii) Neumann conditions on both planes and (iv) mixed conditions, that is, Dirichlet on one plane and Neumann on the other one.   
\end{abstract}

\bigskip

\section{Introduction}

 In 1948, H. B. G. Casimir \cite{Casimir} computed for the first time the energy shift of the vacuum state of QED caused by the presence of two parallel perfectly conducting plates close to each other. For simplicity, and as a first approximation, Casimir considered non-interacting fields\footnote{The expression \lq\lq non-interacting field\rq\rq\ along this paper shall always mean that the field does not interact with itself or with any other field. However, it is sensitive to the presence of material plates, but the interaction between the field and the plates shall be simulated by idealized boundary conditions (BC).} and that the conducting plates influence only the (quantized) electromagnetic field. For this case, the vacuum energy shift can be obtained by the shift in the zero-point energy of the electromagnetic field.  

	Zero-point energy is not a peculiarity of the electromagnetic field, so that any relativistic quantum field shall exhibit a Casimir effect when submitted to boundary conditions (BC). A detailed discussion about many aspects of the Casimir effect can be found in ref's \cite{PlunienMullerGreiner86, MostepanenkoTrunov97, BordagMostepanenkoMohideen2001, Milton, Miltonlivro} and references therein.
	
	However, any meaningful quantum field theory describing nature is constructed by interacting fields. Therefore, in order to study the Casimir effect in its totality we should take into account all orders in perturbation theory. In terms of Feynmann diagrams, we should compute the influence of the BC's not only in the one-loop vacuum bubbles, but also in the two-loop ones, and so on.
	
	Although there is a vast literature concerning one-loop contributions to the Casimir effect\footnote{The most popular method for computing the one-loop Casimir effect was that introduced by Casimir in 1948 \cite{Casimir}, based on the zero point energy of the field. However, many other equivalent techniques are available for this calculation, as for example Green's function method \cite{Miltonlivro}, generalized zeta-function method \cite{Elizaldelivro}, Schwinger's method \cite{Schwinger92, Farina94, Farina94-2, Farina 95} and other methods \cite{BrownMcLay69}.}, the same is not true, comparatively speaking, for two-loop contributions. In the context of QED, two-loop calculations can be found \cite{QED} and for scalar fields in \cite{scalar, Symanzik, KrechDietrichPRA92, NPnossos, nosso1, nosso2}.

	Though the first radiative corrections (two-loop contribution) to the QED Casimir effect does not have an experimental motivation, since it is $\alpha\lambda_{c}/a$ smaller than the one-loop effect, where $\alpha=1/137$, $\lambda_{c}$ is the Compton wavelength of the electron and $a$ is a typical distance in Casimir experiments, it may be relevant in the QCD bag model \cite{QCD}, as pointed out in \cite{Miltonlivro}.
	
	However, there is indeed an important theoretical motivation for studying the Casimir effect at two-loop level, namely. To test renormalizability of QFT under the influence of BC at higher orders in perturbative theory. Idealized BC at higher orders may lead to unsolvable inconsistences. 
	
	In what concerns interacting fields, one of the most popular models of Quantum Field Theory in the literature is the so called $\lambda\phi^4$ self-interaction model. This occurs due to its simplicity and its applicability to describe many physical phenomena, as spontaneous symmetry breaking and Bose-Einstein condensates.
	
	Although the Casimir effect for the scalar field with $\lambda\phi^4$ self-interaction has been studied recently, there are still a few questions about this subject we want to consider here. The first one is the fact that, for some situations involving Neumann BC's, the first radiative correction to the Casimir effect diverges for $d=2$ spatial dimensions in the massive case \cite{NPnossos,nosso1,nosso2}. The second question,  related to the first one, consists in the necessity of introducing renormalization surface counter-terms for the scalar field in some situations to get finite results, as pointed out in the work of reference \cite{Symanzik}.
	
 This paper is devoted to the study of the first radiative corrections to the Casimir energy for the massive complex scalar field $\phi$, with the well known self-interaction $\lambda\phi^{4}$. Specifically, using euclidean coordinates, $(x_{0},x_{1},...,x_{d}=:z)$, in $D=d+1$ dimensions, and a system of units where $\hbar=c=1$, we can write the Lagrangian density of the field in the form
\begin{equation}
\label{Lagrangeanatotal}
{\cal L}_{\rm E}=|\partial_{\mu}\phi|^{2}+m^{2}|\phi|^{2}+\lambda|\phi|^{4}
+{\cal L}_{\rm ct}
\end{equation}
where ${\cal L}_{\rm ct}$ is the lagrangian counter-terms.
We consider the field always submitted to boundary conditions on two parallel planes, located, in our coordinates system, at $x_{d}=z=0$ and $x_{d}=z=a$.

	In fact, the $\lambda\phi^{4}$ theory is renormalizable only for $d=2,3$ spatial dimensions, even when the field is not submitted to any boundary condition. Therefore, our calculations will be valid only in these cases ($d=2,3$).

	This paper has the following structure: in section 2 we develop, up to order $\lambda$, the general quantum theory for the complex massive scalar field described by the Lagrangian density (\ref{Lagrangeanatotal}) and submitted to BC's at two parallel planes (or lines in $d=2$ spatial dimensions) located at $x_{d}=z=0$ and $x_{d}=z=a$. In section 3 we study the case of quasi-periodic BC, which interpolates continuously the periodic and anti-periodic conditions on one spatial direction. In section 4 we study the case where the field obeys Dirichlet BC at both planes. In section 5 we consider the case where the field obeys Neumann BC at both planes, and finally in the section 6 we study the case where the field satisfies Dirichlet BC at $x_{d}=z=0$ and Neumann BC at $x_{d}=z=a$. The last section is devoted to some conclusions and final remarks.

\section{General theory}
	
	Despite the fact that we are considering a field submitted to BC's, its Quantum Theory follows in strict analogy to the case where there is not any BC's. The energy of the vacuum state of the field can be written as \cite{nosso1,Peskin}
\begin{equation}
\label{formaenergia}
E=\lim_{T\rightarrow\infty}-\frac{1}{T}\ln\Biggl[\int{\cal D}\phi{\cal D}\phi^{*}\exp\Biggl(-\int_{-T/2}^{T/2} dx_{0}
\int d^{d}{\bf x}{\cal L}_{E}(x_{0},{\bf x})\Biggr)\Biggr]\ ,
\end{equation}
where ${\bf x}=(x_{1},...,x_{d}=z)$, $x_{0}$ is the temporal coordinate and ${\cal L}_{E}$ is the Lagrangian density (\ref{Lagrangeanatotal}). It is worth emphasizing that we are using Euclidian coordinates.
	
	For all boundary conditions we considered in this paper, the renormalization of the quantum theory described by the Lagrangian (\ref{Lagrangeanatotal}) is assured, in order $\lambda$, by the Lagrangian counter-terms
\begin{equation}
\label{Lcontratermos}
{\cal L}_{ct}=\delta m^{2}|\phi|^{2}-\Bigl(c_{1}\delta(z)+c_{2}\delta(z-a)\Bigr)|\phi|^{2}+\delta\Lambda\ ,
\end{equation}
where $\delta\Lambda$ is a variation in the cosmological constant due to the $\lambda\phi^{4}$ interaction (i.e., the change in the vacuum energy which is due solely to the interaction, and not to the confinement), $\delta m^{2}$ is the mass renormalization counter-term, and $c_{1}$ and $c_{2}$ are surface renormalization counter-terms. These last terms can be interpreted as extra renormalizations for the field mass, which must be taken only at the planes $x_{d}=z=0$ and $x_{d}=z=a$.

	By means of standard perturbative methods of quantum field theory, equation (\ref{formaenergia}) gives the first radiative correction for the Casimir energy per unit of area,
\begin{equation}
\label{energia1geral}
\frac{E^{1}}{A}={\cal E}^{1}=\int_{0}^{a}dz\ \Biggl[2\lambda G_{c}^{2}(x,x)+\delta m^{2}G_{c}(x,x)-\bigl[c_{1}\delta(z)+c_{2}\delta(z-a)\bigr]G_{c}(x,x)+\delta\Lambda\Biggr]\ ,
\end{equation}
with $E^{1}$ representing the first radiative correction for the Casimir energy, $A=\int dx_{1}...\int dx_{d-1}$ is the planes area and $G_{c}(x,y)$ is the Green's function of the field submitted to the boundary conditions, but in the absence of the $\lambda\phi^{4}$ self-interaction term. In writing expression (\ref{energia1geral}) we discarded divergent contributions that come from the integration in the exterior regions of the planes. These contributions do not depend on the distance $a$, and then do not contribute to the Casimir force.    

	In order to calculate the integrals presented in (\ref{energia1geral}), we have first to write the Green's function $G_{c}(x,y)$ in its spectral representation, which can be done by considering the field modes between the planes. Since all boundary conditions considered here are imposed on the $x_{d}=z$ coordinate, these field modes can be written as
\begin{equation}
\label{modosgeral}
\phi(x)=\Phi(x_{\perp})\varphi_{n}(z)\ ,
\end{equation}
where we defined $x_{\perp}=(x_{0},x_{1},...,x_{d-1})$, $\Phi(x_{\perp})$ has the usual expansion for free fields and $\varphi_{n}(z)$ satisfies the BC's. In addition, the functions $\varphi_{n}(z)$ form an orthonormal set according to
\begin{equation}
\label{ortonormalmodos}
\int_{0}^{a}\varphi_{n}(z)\varphi^{*}_{n'}(z)=\delta_{n,n'}\ ,
\end{equation}
with $\delta_{n,n'}$ designating the Kronecker delta.

	Now, the spectral representation for $G_{c}(x,y)$ reads 
\begin{equation}
\label{espcontgeral}
G_{c}(x,x')=\int\frac{d^{d}k_{\perp}}{(2\pi)^{d}}\exp\left[ik_{\perp}\cdot(x_{\perp}-x'_{\perp})\right]\sum_{n}\frac{\varphi_{n}(z)\varphi^{*}_{n}(z')}{k_{\perp}^{2}+q_{n}^{2}+m^{2}}
\end{equation}
where $k_{\perp}=(k_{0},...,k_{d-1})$ and $q_{n}$ is the $d$-th Fourier factor, quantized by the boundary conditions.

	By analytic continuation, equation (\ref{espcontgeral}) gives \cite{nosso1,Kaku}
\begin{equation}
\label{espcontgerl1}
G_{c}(x,x)=\frac{1}{(4\pi)^{d/2}}\Gamma\left(1-\frac{d}{2}\right)\sum_{n}\omega_{n}^{d-2}\varphi_{n}(z)\varphi^{*}_{n}(z)\ ,
\end{equation}
with
\begin{equation}
\label{defomegangeral}
\omega_{n}=\sqrt{q_{n}^{2}+m^{2}}\ .
\end{equation}

	Although the summation written in (\ref{espcontgerl1}) is divergent in the general case, the expression (\ref{espcontgerl1}) shall allow us to calculate the integrals presented in (\ref{energia1geral}) by analytic continuation in all situations considered in this paper.

	The renormalization counter-terms are determinated from the Green's function of the field (with boundary conditions) which, in first order in $\lambda$, reads
\begin{equation}
\label{qwe2}
G(x,y)=G_{c}(x,y)-\int d^{D}\xi G_{c}(x,\xi)\Sigma_{c}(\xi)G_{c}(\xi,y)\ ,
\end{equation}
where
\begin{equation}
\label{Sigmac}
\label{defSigmageral}
\Sigma_{c}(\xi)=4\lambda G_{c}(\xi,\xi)+\delta m^{2}-\bigl[c_{1}\delta(\xi_{d})+c_{2}\delta(\xi_{d}-a)\bigr]
\end{equation}
is the self-energy of the field in the presence of the BC's.

	Although $\delta\Lambda$ does not depend on the boundary condition imposed on the field, it is convenient to calculate this counter-term for each case separately. This is done by imposing $\delta\Lambda$ to be $a$-independent and setting to zero the contribution for the first radiative correction of the Casimir energy which is proportional to the spatial volume between the planes, that is, the contribution linear in the distance $a$ and in the area $A$.

	The mass counter-term $\delta m^{2}$ is calculated by imposing that the quantity $\Sigma_{c}(\xi)$ satisfies the following conditions:
\begin{eqnarray}
\label{condicoesSigma}
\Sigma_{c}(\xi)&\Longrightarrow&\mbox{finite between the planes}\cr
\lim_{a\rightarrow\infty}\Sigma_{c}(\xi)&=&0\ ,
\end{eqnarray}
except, possibly, on the planes. In addition, we consider $\delta m^{2}$ to be $a$-independent.

	From the image method, we can write $G_{c}(x,x')$ in the form \cite{nosso1,nosso2}
\begin{equation}
\label{defGc}
G_{c}(\xi,\xi')=G_{0}(\xi,\xi')+{\tilde G}_{c}(\xi,\xi')
\end{equation}
where $G_{0}(\xi,\xi')$ is the Green's function for the non-interacting field without BC's, and ${\tilde G}_{c}(\xi,\xi')$ is a correction induced by the BC's which is finite in the region between the planes, but can, possibly, diverge on the planes. Besides, ${\tilde G}_{c}(\xi,\xi')$ has the property vanishing when the distance between the planes goes to infinity.

	Substituting (\ref{defGc}) in (\ref{Sigmac}) we can see that we are taken to a divergent term proportional to $G_{0}(\xi,\xi)$. In order to have accordance with the first condition (\ref{condicoesSigma}), this divergence must be canceled. It is done by taking
\begin{equation}
\label{qwe1}
4\lambda G_{0}(\xi,\xi)+\delta m^{2}=\mu^{2}\ ,
\end{equation}
where $\mu^{2}$ is a real finite constant. Substituting (\ref{qwe1}) into (\ref{Sigmac}) we have
\begin{equation}
\Sigma_{c}(\xi)=4\lambda{\tilde G}_{c}(\xi,\xi)+\mu^{2}-\bigl[c_{1}\delta(\xi_{d})+c_{2}\delta(\xi_{d}-a)\bigr]\ .
\end{equation}
Taking the limit $a\rightarrow\infty$ of the above expression and considering the second condition (\ref{condicoesSigma}) we have that $\mu^{2}=0$ (where we used that $\lim_{a\rightarrow\infty}{\tilde G}_{c}(x,y)=0$). Therefore, equation (\ref{qwe1}) yields.
\begin{equation}
\label{deltam2}
\delta m^{2}=-4\lambda G_{0}(\xi,\xi)=-4\lambda\int\frac{d^{d+1}k}{(2\pi)^{(d+1)}}\frac{1}{k^{2}+m^{2}}=-\lambda\frac{4m^{d-1}}{(4\pi)^{(d+1)/2}}\Gamma\Biggl(\frac{1-d}{2}\Biggr)\ ,
\end{equation}
where we used the Furrier representation of $G_{0}(\xi,\xi)$ in euclidean coordinates and calculated the above integral by analytic continuation \cite{Kaku}.

	The mass counter-term expressed in (\ref{deltam2}) is valid for all boundary conditions and dimensions considered in this paper.

	The surface counter-terms $c_{1}$ and $c_{2}$ are determinated by imposing that the first order correction to the Green's function $G(x,y)$ written in (\ref{qwe2}) is finite between the planes. This is equivalent of imposing
\begin{equation}
\label{condicaocorrecaoG}
\int_{0}^{a}d\xi_{d}G_{c}(x,\xi)\Sigma_{c}(\xi)G_{c}(\xi,y)\Longrightarrow\mbox{finite}\ .
\end{equation}
We can not analyze the above expression in a general case. It must be considered for each specific situation separately.

	Now, let us calculate the first radiative correction to the Casimir energy for each BC's mentioned previously.

\section{Quasi-periodic boundary condition}

	In this section we consider the case where the field satisfies the so called quasi-periodic boundary condition, namely,
\begin{equation}
\label{bcQP}
\phi(x_{\perp},x_{d}=z=a)=\exp(i\theta)\phi(x_{\perp},x_{d}=z=0)
\end{equation}
where $\theta$ is a parameter that can vary continuously from $0$ to $\pi$. Note that for $\theta=0$ and $\theta=\pi$ we have, respectively, periodic and anti-periodic conditions.

	With conditions (\ref{bcQP}) the functions $\varphi_{n}(z)$ written in (\ref{modosgeral}) take the form
\begin{equation}
\label{defvarphiQP}
\varphi_{n}(z)=\frac{1}{\sqrt{a}}\exp\left(i\frac{2n\pi+\theta}{a}z\right)\ \ ,\ \ n=0,\pm1,\pm2,...\ \ ,
\end{equation}
with the factors $q_{n}$ appearing in (\ref{espcontgeral}) given by
\begin{equation}
\label{defqnQP}
q_{n}=\frac{2n\pi+\theta}{a}\ .
\end{equation}

	For the BC's (\ref{bcQP}) there is no need to consider the surface counter-terms $c_{1}$ and $c_{2}$ in the Lagrangian density (\ref{Lcontratermos}), because they are not helpful to renormalize any divergence at all\footnote{Recall that imposing BC (\ref{Lcontratermos}) means, in some since, to compactify one spatial dimension. Consequently, we are left to a compact manifold without boundary and hence it is meaningless to talk about surface terms.}. Therefore, equation (\ref{energia1geral}) becomes
\begin{equation}
\label{intE1theta}
{\cal E}^{(1)}_{\theta}=\int_{0}^{a}dx_{d}\left[2\lambda G^{2}_{\theta}(x,x)+\delta m^{2}\,G_{\theta}(x,x)+\delta\Lambda\right]\ ,
\end{equation}
where $G_{\theta}(x,x')$ designates the Green's function for the non-interacting $\phi$ field submitted to the condition (\ref{bcQP}).

	Using equations (\ref{espcontgerl1}), (\ref{defomegangeral}), (\ref{defvarphiQP}) and (\ref{defqnQP}) we can write
\begin {equation}
\label {zxc1}
G_{\theta}(x,x)={\Gamma\left(1-d/2\right)\over(4\pi)^{d/2}a}
\sum_{n=-\infty}^{\infty}\omega_{n}^{d-2}\qquad(d<1),
\end {equation}
with
\begin{equation}
\label{defomegantheta}
\omega_{n}=\sqrt{\left(\frac{2n\pi+\theta}{a}\right)^{2}+m^{2}}\ .
\end{equation}

	Inserting expression (\ref{zxc1}) into equation (\ref{intE1theta}) and arranging terms, we obtain
\begin{equation}
\label{asd}
{\cal E}^{(1)}_{\theta}={2\lambda\over a}\left[{\Gamma(1-d/2)\over(4\pi)^{d/2}}
\sum_{n=-\infty}^{\infty}\omega_{n}^{d-2}
+{a\,\delta m^{2}\over 4\lambda}\right]^{2}+a\left[\delta\Lambda 
-{(\delta m^{2})^{2}\over 8\lambda}\right].
\end{equation}

	In order to compute the summation that appears in equation (\ref{asd}) it is convenient to reexpress it as
\begin{equation}
\label{sum}
\sum_{n=-\infty}^{\infty}\omega_{n}^{d-2}=\left({2\pi\over a}\right)^{d-2}
{\cal D}\left({2-d\over 2},{ma\over 2\pi},{\theta\over 2\pi}\right),
\end{equation}
where we used equation (\ref{defomegantheta}) and defined the function ${\cal D}$ by the expression
\begin{equation}
{\cal D}\left(s,\nu,\frac{\theta}{2\pi}\right):=
\sum_{n=-\infty}^{\infty}\left[\nu^2+\left(n+\frac{\theta}{2\pi}\right)^2
\right]^{-s},\qquad{\rm Re}(s)>1/2.
\end{equation}
This function has the analytic continuation to the whole complex $s$-plane given by \cite{FarinaHenrique}
\begin{equation}
\label{extanalitica}
{\cal D}\left(s,\nu,\frac{\theta}{2\pi}\right)=
\frac{\sqrt{\pi}\,\nu^{1-2s}}{\Gamma(s)}\left[\Gamma\left(s-\frac{1}{2}\right)
+4\sum_{n=1}^{\infty}\cos(n\theta)\,\frac{K_{1/2-s}(2n\pi\nu)}
{(n\pi\nu)^{1/2-s}}\right],
\end{equation}
which exhibits simple poles at $s=1/2,-1/2,-3/2,\ldots$ Inserting equations (\ref{sum}) and (\ref{extanalitica}) into equation (\ref{asd}) we have
\begin{eqnarray}
\label{Eregtheta}
{\cal E}^{(1)}_{\theta}&=&{2\lambda\over a}\left\{\frac{am^{d-1}}{(4\pi)^{(d+1)/2}}
\left[\Gamma\left(\frac{1-d}{2}\right)+4\sum_{n=1}^{\infty}\cos(n\theta)\,
\frac{K_{(d-1)/2}(nma)}{(nma/2)^{(d-1)/2}}\right]
+{a\,\delta m^{2}\over 4\lambda}\right\}^{2}
\nonumber \\
& &+a\left[\delta\Lambda 
-{(\delta m^{2})^{2}\over 2^{3}\lambda}\right].
\end{eqnarray}

	In order to fix $\delta\Lambda$ we require that linear term in $a$ in the above expression of ${\cal E}^{1}_{\theta}$ must vanish, as explained before. So we have
\begin{equation}
\label{dL}
\delta\Lambda=\frac{(\delta m^2)^2}{2^{3}\lambda}\ .
\end{equation}

	Inserting (\ref{deltam2}) and (\ref{dL}) into equation (\ref{Eregtheta}) we finally arrive at the first radiative correction to the Casimir energy per unity area for quasi-periodic boundary conditions:
\begin{equation}
\label{finaltheta}
{\cal E}^{(1)}_{\theta}(a)=\frac{\lambda m^{d-1}}{2^{d-2}\pi^{d+1}a^{d-2}}
\left[\sum_{n=1}^{\infty}\cos(n\theta)
{K_{(d-1)/2}(man)\over n^{(d-1)/2}}\right]^{2}.
\end{equation}

	As mentioned before, expression (\ref{finaltheta}) is meaningful only for $d=3,2$ spatial dimensions;
\begin{equation}
\label{final3dtheta}
{\cal E}^{(1)}_{\theta}(a)\Big|_{d=3}=\frac{\lambda m^2}{2\pi^4 a}
\left[\sum_{n=1}^{\infty}\cos(n\theta)\,\frac{K_1(nma)}{n}\right]^2\ ,
\end{equation}
\begin{equation}
\label{final2dtheta}
{\cal E}^{(1)}_{\theta}(a)\Big|_{d=2}= \frac{2\lambda}{a}\frac{1}{(4\pi)^{2}}\Bigl[\ln\Bigl(1+e^{-2ma}-2e^{-ma}\cos(\theta)\Bigr)\Bigr]^{2} \ ,
\end{equation}
where we used expression (\ref{somaparatheta}).

	 For the special cases of periodic ($\theta=0$) and anti-periodic ($\theta=\pi$) BC's, equations (\ref{finaltheta}) reduce to
\begin{eqnarray}
\label{finaldP}
{\cal E}^{(1)}_{P}(a)&=&{\cal E}^{(1)}_{\theta=0}(a)=\frac{\lambda m^{d-1}}{2^{d-2}\pi^{d+1}a^{d-2}}\left[\sum_{n=1}^{\infty}{K_{(d-1)/2}(man)\over n^{(d-1)/2}}\right]^{2}\ ,\\
\label{finaldA}
{\cal E}^{(1)}_{A}(a)&=&{\cal E}^{(1)}_{\theta=\pi}(a)=\frac{\lambda m^{d-1}}{2^{2d-5}\pi^{d+1}a^{d-2}}\Biggl[\sum_{n=1}^{\infty}\frac{K_{(d-1)/2}(2man)}{(man)^{(d-1)/2}}\cr\cr
&\ &\ \ \ \ \  \ \ \ \ \ \ \ \ \ \ \ -2^{(d+1)/2}\sum_{n=1}^{\infty}\frac{K_{(d-1)/2}(man)}{(man)^{(d-1)/2}}\Biggr]^{2}\ .
\end{eqnarray}
Similarly, equations (\ref{final3dtheta}) and (\ref{final2dtheta}) take the form
\begin{eqnarray}
\label{final3P}
{\cal E}^{(1)}_{P}(a)\Big|_{d=3}&=&{\cal E}^{(1)}_{\theta=0}(a)\Big|_{d=3}=\frac{\lambda m^{2}}{2\pi^{4}a}\Biggl[\sum_{n=1}^{\infty}\frac{K_{1}(man)}{n}\Biggr]^{2}\ , \\
\label{final3A}
{\cal E}^{(1)}_{A}(a)\Big|_{d=3}&=&{\cal E}^{(1)}_{\theta=\pi}(a)\Big|_{d=3}=\frac{\lambda m^{2}}{2\pi^{4}a}\Biggl[\sum_{n=1}^{\infty}\frac{K_{1}(2man)-K_{1}(man)}{n}\Biggr]^{2}\ ,\\
\label{final2P}
{\cal E}^{(1)}_{P}(a)\Big|_{d=2}&=&{\cal E}^{(1)}_{\theta=0}(a)\Big|_{d=2}=\frac{\lambda}{2\pi^{2}a}\Bigl[\ln\Bigl(1-e^{-ma}\Bigr)\Bigr]^{2}\ ,\\
\label{final2A}	 
{\cal E}^{(1)}_{A}(a)\Big|_{d=2}&=&{\cal E}^{(1)}_{\theta=\pi}(a)\Big|_{d=2}=\frac{\lambda}{2\pi^{2}a}\Bigl[\ln\Bigl(1+e^{-ma}\Bigr)\Bigr]^{2}\ .
\end{eqnarray}

	If we further take the limit $m\to 0$ in equations (\ref{finaltheta}), (\ref{final3dtheta}) and (\ref{final2dtheta}) we have
\begin{eqnarray}
\label{thetam=0,d}
{\cal E}^{(1)}_{\theta}(a)\Bigl|_{m=0}&=&\frac{\lambda}{2\pi^{d+1}a^{2d-3}}\Gamma^{2}\Biggl(\frac{d-1}{2}\Biggr)\Biggl(\sum_{n=1}^{\infty}\frac{\cos(n\theta)}{n^{d-1}}\Biggr)^{2}\ ,\\
\label{thetam=0,d=3}
{\cal E}^{(1)}_{\theta}(a)\Big|_{d=3\atop m=0}&=&\frac{\lambda}{2\pi^4 a^3}
\left[\sum_{n=1}^{\infty}\frac{\cos(n\theta)}{n^2}\right]^2
=\frac{\lambda}{2a^3}\left[B_2\left(\frac{\theta}{2\pi}\right)\right]^2\ ,\\
\label{thetam=0,d=2}
{\cal E}^{(1)}_{\theta}(a)\Big|_{d=2\atop m=0}&=&\frac{\lambda}{2\pi^{2}a}\Biggl[\ln\Biggl(2\sin\Bigl(\theta/2\Bigr)\Biggr)\Biggr]^{2}\ ,
\end{eqnarray}
where we used equations (\ref{zxcap}) and (\ref{zxcap1}), and $B_2(x)=x^2-x+1/6$ is the Bernoulli polynomial of second degree \cite{GR}.

	For the specific cases of periodic ($\theta=0$) and anti-periodic ($\theta=\pi$) conditions, equation (\ref{thetam=0,d}) leads to.
\begin{eqnarray}
\label{Pm=0,d}
{\cal E}_{P}^{(1)}(a)\Big|_{m=0}={\cal E}_{\theta=0}^{(1)}(a)\Big|_{m=0}
&=&\frac{\lambda}{2\pi^{d+1}a^{2d-3}}\Gamma^{2}\Biggl(\frac{d-1}{2}\Biggr)\zeta^{2}(d-1)\ ,\\
\label{Am=0,d}
{\cal E}_{A}^{(1)}(a)\Big|_{m=0}={\cal E}_{\theta=\pi}^{(1)}(a)\Big|_{m=0}
&=&\frac{\lambda}{2\pi^{d+1}a^{2d-3}}\Biggl(\frac{1}{2^{d-2}}-1\Biggr)^{2}\Gamma^{2}\Biggl(\frac{d-1}{2}\Biggr)\zeta^{2}(d-1)\ .
\end{eqnarray}

	The above results for $d=3$ become
\begin{eqnarray}
\label{Pm=0,d=3}
{\cal E}_{P}^{(1)}(a)\Big|_{d=3\atop m=0}={\cal E}_{\theta=0}^{(1)}(a)\Big|_{d=3\atop m=0}&=&\frac{1}{2^{3}3^{2}a^{3}}\ ,\\
\label{Am=0,d=3}
{\cal E}_{A}^{(1)}(a)\Big|_{d=3\atop m=0}={\cal E}_{\theta=\pi}^{(1)}(a)\Big|_{d=3\atop m=0}&=&\frac{1}{2^{5}3^{2}a^{3}}\ ,
\end{eqnarray}
which could be obtained from expression (\ref{thetam=0,d=3}). Similarly, for $d=2$, we get
\begin{eqnarray}
\label{Pm=0,d=2}
{\cal E}_{P}^{(1)}(a)\Big|_{d=2\atop m=0}={\cal E}_{\theta=0}^{(1)}(a)\Big|_{d=2\atop m=0}&\rightarrow&\infty\ ,\\
\label{Am=0,d=2}
{\cal E}_{A}^{(1)}(a)\Big|_{d=2\atop m=0}={\cal E}_{\theta=\pi}^{(1)}(a)\Big|_{d=2\atop m=0}&=&\frac{\lambda}{2\pi^{2}a}\Bigl[\ln(2)\Bigr]^{2}\ .
\end{eqnarray}
which could also be calculated using expression (\ref{thetam=0,d=2}).

	The above results are in perfect agreement with those presented in the literature \cite{KrechDietrichPRA92}. It is worthwhile to emphasize that for $m=0$ and $d=2$ the first radiative correction to the Casimir energy per unity area diverges in the periodic case ($\theta=0$), as expressed in equation (\ref{Pm=0,d=2}). A similar situation occurs for the field at finite temperature: there, for $d=2$ spatial dimensions, we have a divergence when $m=0$.

	From result (\ref{finaltheta}) we can see that the first radiative correction to the Casimir energy per unity area for quasi-periodic boundary conditions is always positive, whatever parameter $\theta$ we use.

\section{Dirichlet boundary conditions - DD}

	In this section we consider the field submitted to Dirichlet boundary conditions at both planes, it is
\begin{equation}
\label{bcDD}
\phi(x_{\perp},x_{d}=z=0)=\phi(x_{\perp},x_{d}=z=a)=0\ ,
\end{equation}
a situation which, from now on, we shall refer to as DD conditions. In this case, the functions $\varphi_{n}$ appearing in (\ref{modosgeral}) are given by
\begin{equation}
\label{defvarphinDD}
\varphi_{n}(z)=\sqrt{\frac{2}{a}}\sin\left(\frac{n\pi z}{a}\right)\ \ ,\ \ n=1,2,...\ ,
\end{equation}
and the $q_{n}$ factors become
\begin{equation}
q_{n}=\frac{n\pi}{a}\ .
\end{equation}
As a consequence, equation (\ref{defomegangeral}) gives
\begin{equation}
\label{omeganDD}
\omega_{n}=\sqrt{\left(\frac{n\pi}{a}\right)^{2}+m^{2}}
\end{equation}

	Once the field satisfies DD conditions, the corresponding Green's function $G_{DD}$ shall, likewise, satisfy (\ref{bcDD}). So, their contributions to the integral in (\ref{energia1geral}) vanish, whatever values of $c_{1}$ and $c_{2}$ we use. Therefore the counter-terms $c_{1}$ and $c_{2}$ can be set to zero and equation (\ref{energia1geral}) becomes
\begin{equation}
\label{intE1DD}
{\cal E}^{(1)}_{DD}=\int_{0}^{a}d\xi_{d}\left[2\lambda G^{2}_{DD}(\xi,\xi)+\delta m^{2}\,G_{DD}(\xi,\xi)+\delta\Lambda\right]\ ,
\end{equation}
where $G_{DD}(\xi,\xi')$ is the Green's function for the non-interacting field submitted to the DD conditions. Besides, taking into account the Lagrangian counter-terms (\ref{Lcontratermos}), we can see that the contributions of the counter-terms $c_{1}$ and $c_{2}$ for the action (integral of (\ref{Lagrangeanatotal})) is always zero, due to the condition (\ref{bcDD}).

	Using equations (\ref{espcontgerl1}) and (\ref{defvarphinDD}), we can write the Casimir energy (\ref{intE1DD}) in the form
\begin{eqnarray}
{\cal E}^{1}_{DD}&=&\frac{2\lambda}{a}\frac{\Gamma^{2}(1-d/2)}{(4\pi)^{d}}\Biggl[\Biggr(\sum_{n=1}^{\infty}\omega_{n}^{d-2}\Biggl)^{2}+\frac{1}{2}\sum_{n=1}^{\infty}\omega_{n}^{2d-4}\Biggl]\cr\cr
&+&\delta m^{2}\frac{\Gamma\left(1-(d/2)\right)}{(4\pi)^{d/2}}\sum_{n=1}^{\infty}\omega_{n}^{d-2}+a\delta\Lambda\ .
\end{eqnarray}
Completing the square, we have
\begin{eqnarray}
\label{DDintermediario1}
{\cal E}^{1}_{DD}&=&\frac{2\lambda}{a}\Biggl[\Biggl(\frac{\Gamma(1-d/2)}{(4\pi)^{d/2}}\sum_{n=1}^{\infty}\omega_{n}^{d-2}+\frac{\delta m^{2} a}{4\lambda}\Biggr)^{2}+\frac{\Gamma^{2}(1-d/2)}{2(4\pi)^{d}}\sum_{n=1}^{\infty}\omega_{n}^{2d-4}\Biggr]\cr\cr
&+&a\Biggl(\delta\Lambda-\frac{(\delta m^{2})^{2}}{2^{3}\lambda}\Biggr)\ .
\end{eqnarray}

	By the same arguments presented for the quasi-periodic BC's, the counter term $\delta\Lambda$ for DD conditions is also given by (\ref{dL}). The summations appearing in (\ref{DDintermediario1}) can be calculated by analytic continuation (see the Appendix). Therefore, using equations (\ref{deltam2}), (\ref{dL}), (\ref{omeganDD}), (\ref{DDintermediario1}) and (\ref{extanalitica2}), we have the first radiative correction for the Casimir energy per unity area with the field submitted to DD BC's in $d$ spatial dimensions:
\begin{eqnarray}
\label{E1DD}
{\cal E}^{1}_{DD}&=&\frac{2\lambda}{a}\Biggl[\Biggr(\frac{4am^{d-1}}{(4\pi)^{(d+1)/2}}\sum_{n=1}^{\infty}\frac{K_{(d-1)/2}(2man)}{(man)^{(d-1)/2}}\Biggr)^{2}+\cr\cr
&-&2\frac{4am^{d-1}}{(4\pi)^{(d+1)/2}}\frac{\Gamma(1-d/2)}{2(4\pi)^{d/2}}m^{d-2}\sum_{n=1}^{\infty}\frac{K_{(d-1)/2}(2man)}{(man)^{(d-1)/2}}+\cr\cr
&+&\frac{2am^{2d-3}}{(4\pi)^{d+1/2}}\frac{\Gamma^{2}(1-d/2)}{\Gamma(2-d)}\sum_{n=1}^{\infty}\frac{K_{(2d-3)/2}(2man)}{(man)^{(2d-3)/2}}\Biggr]\ ,
\end{eqnarray}
where we discarded an $a$-independent term which does not contribute to the Casimir force.

	For $d=3,2$ spatial dimensions, we have, respectively
\begin{eqnarray}
\label{E1DDd=3}
{\cal E}^{1}_{DD}\Bigl|_{d=3}&=&\frac{\lambda m^{2}}{32\pi^{2}a}\Biggr[\Biggl(1+\frac{2}{\pi}\sum_{n=1}^{\infty}\frac{K_{1}(2man)}{n}\Biggl)^{2}-1\Biggr]\ ,\\
\label{E1DDd=2}
{\cal E}^{1}_{DD}\Bigl|_{d=2}&=&\frac{\lambda\gamma^{2}}{32\pi^{2}a}\Biggl[\Biggl(1+\frac{2}{\gamma}\ln\Bigl(1-\exp(-2ma)\Bigr)\Biggr)^{2}-1\Biggr]\ ,
\end{eqnarray}
where we used the fact that $\Gamma(-1/2)=-2\sqrt{\pi}$ and $\Gamma(-1)\to\infty$ for the result (\ref{E1DDd=3}). For the result (\ref{E1DDd=2}) we used equation (\ref{qweap1}) and the limit
\begin{equation}
\label{rty}
\lim_{d\rightarrow2}\Biggl[\Gamma\Bigl(1-d/2\Bigr)\Biggl(\frac{\Gamma\Bigl(1-d/2\Bigr)}{\Gamma\Bigl(2-d\Bigr)}-2\Biggr)\Biggr]=2\gamma\ ,
\end{equation}
with $\gamma$ being the Euler's constant.

	For a vanishing mass, the limit $m\rightarrow 0$ in equation (\ref{E1DD}) yields
\begin{eqnarray}
\label{ert1}
{\cal E}^{1}_{DD}\Big|_{m=0}&=&2\lambda\frac{a^{3-2d}}{(4\pi)^{d+1/2}}\Biggl[\frac{4}{(4\pi)^{1/2}}\Gamma^{2}\Biggl(\frac{d-1}{2}\Biggr)\zeta^{2}(d-1)+\cr\cr
&+&\Gamma^{2}\Biggl(\frac{2-d}{2}\Biggr)\Gamma\Biggl(\frac{2d-3}{2}\Biggr)\frac{\zeta(2d-3)}{\Gamma(2-d)}\Biggr]\ ,
\end{eqnarray}
where we considered that $\lim_{m\rightarrow0}m^{d-2}=0$, used equation (\ref{zxcap}) and $\zeta$ designates the Riemann zeta function.	Taking the limits $d\to3,2$ of equation (\ref{ert1}), we have, respectively
\begin{eqnarray}
\label{DDm=03}
{\cal E}^{1}_{DD}\Big|_{d=3,m=0}&=&\frac{\lambda}{2^{7}3^{2}a^{3}}\ ,\\
\label{DDm=02}
{\cal E}^{1}_{DD}\Big|_{d=2,m=0}&=&-\frac{\lambda}{2^{5}3a}\Biggl[1-\frac{12}{\pi^{2}}\Bigl(2\gamma_{1}+\gamma^{2}\Bigr)\Biggr]\ ,
\end{eqnarray}
with
\begin{equation}
\label{rty3}
\gamma_{1}=\frac{1}{2}\frac{\partial^{2}}{\partial s^{2}}\Bigl[(1-s)\zeta(s)\Bigr]\Big|_{s=1}\ .
\end{equation}

	Results (\ref{DDm=03}) and (\ref{DDm=02}) agree with those presented in the literature for $m=0$ \cite{Symanzik,KrechDietrichPRA92}. Expression (\ref{DDm=03}) could be obtained from (\ref{E1DDd=3}) in the limit $m\to0$ and with the aid of (\ref{ert2}). The limit $m\to0$ of equation (\ref{E1DDd=2}) is divergent. For this case ($m=0$ and $d=2$) we have to take first $m\to0$ considering $\lim_{m\rightarrow0}m^{d-2}=0$, and after that we take $d\to2$, as we have made for computing (\ref{DDm=02}).

\section{Neumann boundary conditions - NN}

	For this case, the field satisfy the following conditions
\begin{equation}
\label{dfgNN}
\frac{\partial\phi(x)}{\partial x}\Bigg{|}_{x_{d}=z=0}=\frac{\partial\phi(x)}{\partial x}\Bigg{|}_{x_{d}=z=a}=0\ .
\end{equation}
With the restriction (\ref{dfgNN}) the functions $\varphi_{n}$ appearing in (\ref{modosgeral}) read
\begin{equation}
\label{varphiNN}
\varphi_{n}(z)=\sqrt{\frac{2-\delta_{n,0}}{a}}\cos\Biggl(\frac{n\pi z}{a}\Biggr)\ \ \ , \ \ \ n=0,1,2,3,...\ ,
\end{equation}
with the parameter $q_{n}$ and frequencies $\omega_{n}$, defined respectively in (\ref{espcontgeral}) and (\ref{defomegangeral}), given by
\begin{equation}
\label{qnNN,omeganNN}
q_{n}=\frac{n\pi}{a}\ \ ,\ \ \omega_{n}=\sqrt{\frac{n\pi}{a}+m^{2}}\ .
\end{equation}

	For NN conditions the first radiative correction for the Casimir energy per unity area (\ref{energia1geral}) becomes
\begin{equation}
\label{intE1NN}
{\cal E}^{1}_{NN}=\int_{0}^{a}d\xi_{d}\left[ 2\lambda G_{NN}^{2}(\xi,\xi)+\delta m^{2}G_{NN}(\xi,\xi)-c\bigl[\delta(\xi_{d})+\delta(\xi_{d}-a)\bigr]G_{NN}(\xi,\xi)+\delta\Lambda\right]\ ,
\end{equation}
where we used the fact that $c_{1}=c_{2}$ due to the symmetry of the system, and defined $c:=c_{1}=c_{2}$.

	Using equations (\ref{espcontgerl1}) and (\ref{varphiNN}), expression (\ref{intE1NN}) takes the for
\begin{eqnarray}
\label{asd1}
{\cal E}^{1}_{NN}&=&\frac{2\lambda}{a}\frac{\Gamma^{2}\left(1-(d/2)\right)}{(4\pi)^{d}}\left[\left(\sum_{n=0}^{\infty}\omega_{n}^{d-2}\right)^{2}+\frac{1}{2}\sum_{n=1}^{\infty}\omega_{n}^{2d-4}\right]+\cr\cr
&+&\delta m^{2}\frac{\Gamma\left(1-(d/2)\right)}{(4\pi)^{d/2}}\sum_{n=0}^{\infty}\omega_{n}^{d-2}+\cr\cr
&-&\frac{c}{a}\frac{\Gamma\left(1-(d/2)\right)}{(4\pi)^{d/2}}\sum_{n=0}^{\infty}\left(2-\delta_{n,0}\right)\omega_{n}^{d-2}+a\delta\Lambda\ .
\end{eqnarray}

	Completing square and recalling that $\omega_{0}=m$, what can be seen from (\ref{qnNN,omeganNN}), we write equation (\ref{asd1}) in the form
\begin{eqnarray}
\label{qwe}
{\cal E}^{1}_{NN}&=&\frac{2\lambda}{a}\Biggl[\left(\frac{\Gamma\left(1-(d/2)\right)}{(4\pi)^{d/2}}\sum_{n=0}^{\infty}\omega_{n}^{d-2}+\frac{\delta m^{2}a}{4\lambda}\right)^{2}+\cr\cr
&+&\frac{\Gamma^{2}(1-d/2)}{2(4\pi)^{d}}\sum_{n=1}^{\infty}\omega_{n}^{2d-4}+\frac{c}{\lambda}\frac{\Gamma\left(1-d/2\right)}{2(4\pi)^{d/2}}\Biggl(m^{d-2}-2 \sum_{n=0}^{\infty}\omega_{n}^{d-2}\Biggr)\Biggr]+\cr\cr
&+&a\left(\delta\Lambda-\frac{(\delta m^{2})^{2}}{2^{3}\lambda}\right)\ .
\end{eqnarray}

	In order to calculate the surface counter-term $c$, we consider condition (\ref{condicaocorrecaoG}) and write the self-energy $\Sigma_{NN}(\xi)$ expressed in (\ref{defSigmageral}) for the NN conditions as,
\begin{equation}
\label{defSigmaNN}
\Sigma_{NN}(\xi)=4\lambda G_{NN}(\xi,\xi)+\delta m^{2}-c\bigl[\delta(\xi_{d})+\delta(\xi_{d}-a)\bigr]\ .
\end{equation}
In this case, integral (\ref{condicaocorrecaoG}) diverges due to a divergent contribution which comes from $\Sigma_{NN}(\xi)$ and is given by the Green's function evaluated in the coincident point limit and also lying at one of the planes, that is, $G_{NN}(\xi,\xi)$ with $\xi_{d}=z=0$ or $\xi_{d}=a$. This divergence of the integral (\ref{condicaocorrecaoG}) coming from 
$\Sigma_{NN}(\xi)$, expressed in (\ref{defSigmaNN}), can not be renormalized by the mass counter-term $\delta m^{2}$.

	Therefore, the divergence of the integral (\ref{condicaocorrecaoG}) comes from the integration over regions near the planes, and condition (\ref{condicaocorrecaoG}) is equivalent of imposing
\begin{equation}
\label{zxcNN1}
\int_{0}^{\varepsilon}d\xi_{d}G_{NN}(x,\xi)\Sigma_{NN}(\xi)G_{NN}(\xi,y)+\int_{a-\varepsilon}^{a}d\xi_{d}G_{NN}(x,\xi)\Sigma_{NN}(\xi)G_{NN}(\xi,y)\rightarrow\ finite\ ,
\end{equation}
for an arbitrary $\varepsilon$, since $\varepsilon<a/2$, and for $x,y\not=\xi$.

		Using the fact that
\begin{equation}
\frac{\partial G_{NN}(x,\xi)}{\partial\xi_{d}}\Bigg|_{\xi_{d}=0,a\ ; \ \xi_{d}\not=x_{d}=z}=0\ ,
\end{equation}
we see that $G_{NN}(x,\xi)$ does not depend on the coordinate $\xi_{d}$ near the planes. Therefore, for a small $\varepsilon$, equation (\ref{zxcNN1}) can be written as
\begin{eqnarray}
G_{NN}(x,\xi)\big|_{\xi_{d}=0}\Bigl[\int_{0}^{\varepsilon}d\xi_{d}\Sigma_{NN}(\xi,\xi)\Bigr]G_{NN}(\xi,y)\big|_{\xi_{d}=0}\cr\cr
+\ G_{NN}(x,\xi)\big|_{\xi_{d}=a}\Bigl[\int_{a-\varepsilon}^{a}d\xi_{d}\Sigma_{NN}(\xi,\xi)\Bigr]G_{NN}(\xi,y)\big|_{\xi_{d}=a}\rightarrow\ finite\ .
\end{eqnarray}
Once $G_{NN}(x,\xi)\big|_{\xi_{d}=0,a}$ is finite (for $\xi_{d}\not=x^{d}=z$), we have that
\begin{equation}
\label{zxcNN2}
\int_{0}^{\varepsilon}d\xi_{d}\Sigma_{NN}(\xi,\xi)+\int_{a-\varepsilon}^{a}d\xi_{d}\Sigma_{NN}(\xi,\xi)\rightarrow\ finite\ .
\end{equation}
Adding the finite contribution $\int_{\varepsilon}^{a-\varepsilon}\Sigma_{NN}(\xi,\xi)$ to both sides of (\ref{zxcNN2}), we conclude that condition (\ref{condicaocorrecaoG}), for the case at hand, can be substituted by he following one: 
\begin{equation}
\label{zxcNN3}
\int_{0}^{a}\Sigma_{NN}(\xi,\xi)\rightarrow\ finite\ .
\end{equation}

	Substituting (\ref{defSigmaNN}) into (\ref{zxcNN3}) and using expressions (\ref{espcontgerl1}), (\ref{varphiNN}) and (\ref{qnNN,omeganNN}) we obtain
\begin{equation}
\label{zxcNN4}
4\lambda\frac{\Gamma\Bigl(1-(d/2)\Bigr)}{(4\pi)^{d/2}}\sum_{n=0}^{\infty}\omega_{n}^{d-2}+\delta m^{2}a-c\rightarrow\ finite\ ,
\end{equation}
where we considered $\int_{0}^{a}\delta(z^{d})=\int_{0}^{a}\delta(z^{d}-a)=1/2$.

	Taking into account the definition of $\omega_{n}$, expressed in (\ref{qnNN,omeganNN}), the analytic extension (\ref{extanalitica2}) and the result (\ref{deltam2}), condition (\ref{zxcNN4}) can be written in the form
\begin{equation}
\label{zxcNN5}
\left[2\lambda\frac{\Gamma\Bigl(1-(d/2)\Bigr)}{(4\pi)^{d/2}}m^{d-2}-c\right]+\frac{16\lambda}{(4\pi)^{(d+1)/2}}am^{d-1}\sum_{n=1}^{\infty}\frac{K_{(d-1)/2}(2man)}{(man)^{(d-1)/2}} \rightarrow\ finite\ .
\end{equation}

	Due to the presence of the $\Gamma$ function, the first term on the right hand side of (\ref{zxcNN5}) produces a divergent $a$-independent contribution for $d=2$. This problem can be circumvented by setting
\begin{equation}
\label{zxcNN6}
c=2\lambda\frac{\Gamma\Bigl(1-(d/2)\Bigr)}{(4\pi)^{d/2}}m^{d-2}-\lambda m^{d-2}\sigma_{NN}(d)\ ,
\end{equation}
where $\sigma_{NN}(d)$ a dimensionless function of the dimension $d$ and the product $m^{d-2}\sigma_{NN}(d)$ is finite for any $d$.

	In contrast, for $d=3$, expression (\ref{zxcNN5}) does not contain any divergence and the use of the surface counter-term $c$ is not necessary. Therefore, $\sigma_{NN}(d)$ introduced in equation (\ref{zxcNN6}) is chosen such that $c=0$ for $d=3$:
\begin{equation}
\label{sigma(3)}
\sigma(d=3)=2\frac{\Gamma\Bigl(1-(3/2)\Bigr)}{(4\pi)^{3/2}}=-\frac{1}{2\pi}\ .
\end{equation}
However, this is not sufficient to fix $\sigma_{NN}(d)$ completely. Hence, for $d=2$, the value $\sigma(d=2)$ can not be determined.

	The last counter-term appearing in (\ref{qwe}) is the cosmological constant counter-term $\delta\Lambda$. It is given by (\ref{dL}) and determined by the same arguments used for the boundary conditions considered previously.
	
	Substituting (\ref{deltam2}), (\ref{dL}) and (\ref{zxcNN6}) in (\ref{qwe}), using the definitions (\ref{qnNN,omeganNN}) and the analytic extensions (\ref{extanalitica2}) and (\ref{extanalitica2}), discarding an $a$-independent term which does not contribute to the Casimir force and collecting terms, we have the first radiative correction to the Casimir energy per unity area for NN conditions, namely,
\begin{eqnarray}
\label{E1NN}
{\cal E}^{1}_{NN}&=&\frac{2\lambda}{a}\Biggl[\Biggr(\frac{4am^{d-1}}{(4\pi)^{(d+1)/2}}\sum_{n=1}^{\infty}\frac{K_{(d-1)/2}(2man)}{(man)^{(d-1)/2}}\Biggr)^{2}+\cr\cr
&-&2\frac{4am^{d-1}}{(4\pi)^{(d+1)/2}}\left(\frac{\Gamma(1-d/2)}{2(4\pi)^{d/2}}-\frac{\sigma(d)}{2}\right)m^{d-2}\sum_{n=1}^{\infty}\frac{K_{(d-1)/2}(2man)}{(man)^{(d-1)/2}}+\cr\cr
&+&\frac{2am^{2d-3}}{(4\pi)^{d+1/2}}\frac{\Gamma^{2}(1-d/2)}{\Gamma(2-d)}\sum_{n=1}^{\infty}\frac{K_{(2d-3)/2}(2man)}{(man)^{(2d-3)/2}}\Biggr]
\end{eqnarray}

	Taking $d=2$ and $d=3$, which are the relevant cases, we obtain
\begin{eqnarray}
\label{E1NNd=3}
{\cal E}^{1}_{NN}\Bigl|_{d=3}&=&\frac{\lambda m^{2}}{32\pi^{2}a}\Biggr[\Biggl(1-\frac{2}{\pi}\sum_{n=1}^{\infty}\frac{K_{1}(2man)}{n}\Biggl)^{2}-1\Biggr]\ ,\\
\label{E1NNd=2}
{\cal E}^{1}_{NN}\Bigl|_{d=2}&=&\frac{\lambda\gamma^{2}}{32\pi^{2}a}\left[\left[\left(1+\frac{4\pi\sigma(2)}{\gamma}\right)+\frac{2}{\gamma}\ln\Bigl(1-\exp(-2ma)\Bigr)\right]^{2}-\left(1+\frac{4\pi\sigma(2)}{\gamma}\right)^{2}\right]\ ,
\end{eqnarray}
where we considered that $\Gamma(-1/2)=-2\sqrt{\pi}$, $\Gamma(-1)\to\infty$ and used equations (\ref{sigma(3)}), (\ref{rty}) and (\ref{qweap1}).

	The zero mass limits are obtained by taking $m\to0$ in equation (\ref{E1NN}). The calculations are performed analogously to those made for the DD case (\ref{ert1}) and the result is given by
\begin{eqnarray}
\label{rty2}
{\cal E}^{1}_{NN}\Big|_{m=0}={\cal E}^{1}_{DD}\Big|_{m=0}&=&2\lambda\frac{a^{3-2d}}{(4\pi)^{d+1/2}}\Biggl[\frac{4}{(4\pi)^{1/2}}\Gamma^{2}\Biggl(\frac{d-1}{2}\Biggr)\zeta^{2}(d-1)+\cr\cr
&+&\Gamma^{2}\Biggl(\frac{2-d}{2}\Biggr)\Gamma\Biggl(\frac{2d-3}{2}\Biggr)\frac{\zeta(2d-3)}{\Gamma(2-d)}\Biggr]\ .
\end{eqnarray}

	For $d=2,3$ we have
\begin{eqnarray}
\label{NNm=03}
{\cal E}^{1}_{NN}\Big|_{d=3,m=0}&=&\frac{\lambda}{2^{7}3^{2}a^{3}}\ ,\\
\label{NNm=02}
{\cal E}^{1}_{NN}\Big|_{d=2,m=0}&=&-\frac{\lambda}{2^{5}3a}\Biggl[1-\frac{12}{\pi^{2}}\Bigl(2\gamma_{1}+\gamma^{2}\Bigr)\Biggr]
\end{eqnarray}
where $\gamma_{1}$ is defined in (\ref{rty3}) and $\gamma$, as usual, the Euler's constant. The results (\ref{NNm=03}) and (\ref{NNm=02}) are in perfec agreement with those found in the literature \cite{Symanzik,KrechDietrichPRA92}.

	The result (\ref{NNm=03}) could be obtained by taking the limit $m\rightarrow0$ in (\ref{E1NNd=3}).	However, if we take this same limit in (\ref{E1NNd=2}) we get a divergent result, in contrast with the finite result given by (\ref{NNm=02}). The reasons for such a disagreement are the same as those presented for the DD case, where an identical situation occurs.

\section{Mixed boundary conditions - DN}

	In this section we consider the field satisfying Dirichlet condition at $z=0$ and Neumann condition at $z=a$:
\begin{equation}
\label{bcDN}
\phi(x_{\perp},x_{d}=z=0)=\frac{\partial\phi(x)}{\partial x}\Bigg{|}_{x_{d}=z=a}=0\ ,
\end{equation}
which restrict the $\varphi_{n}$ functions written in (\ref{modosgeral}) to be given by
\begin{equation}
\label{varphiDN}
\varphi_{n}(z)=\sqrt{\frac{2}{a}}\sin\left[\left(n+\frac{1}{2}\right)\frac{\pi z}{a}\right]\ \ \ , \ \ \ n=0,1,2,3,...\ .
\end{equation}
Also, $q_{n}$ and $\omega_{n}$, defined respectively in (\ref{espcontgeral}) and (\ref{defomegangeral}), take the form
\begin{equation}
\label{qn;omegaDN}
q_{n}=\left(n+\frac{1}{2}\right)\frac{\pi}{a}\ \ \ , \ \ \ \omega_{n}=\sqrt{\left[\left(n+\frac{1}{2}\right)\frac{\pi}{a}\right]^{2}+m^{2}}
\end{equation}

	By the same reasons as those used for the DD conditions, the counter-term $c_{1}$ appearing in expression (\ref{energia1geral}) must be zero. As a consequence, the first correction to the Casimir energy (\ref{energia1geral}) is written in the form
\begin{equation}
\label{rty4}
{\cal E}^{1}_{DN}=\int_{0}^{a}d\xi_{d}\left[2\lambda G_{DN}^{2}(\xi,\xi)+\delta m^{2}G_{DN}(\xi,\xi)-c\delta(\xi_{d}-a)G_{DN}(\xi,\xi)+\delta\Lambda\right]\ ,
\end{equation}
where, for simplicity, we have done $c=c_{2}$.

	Using equations (\ref{espcontgerl1}) and (\ref{varphiDN}), the integral (\ref{rty4}) yields
\begin{eqnarray}
\label{poi3}
{\cal E}^{1}_{DN}&=&\frac{2\lambda}{a}\frac{\Gamma^{2}\left(1-(d/2)\right)}{(4\pi)^{d}}\left[\left(\sum_{n=0}^{\infty}\omega_{n}^{d-2}\right)^{2}+\frac{1}{2}\sum_{n=0}^{\infty}\omega_{n}^{2d-4}\right]+\cr\cr
&+&\delta m^{2}\frac{\Gamma\left(1-(d/2)\right)}{(4\pi)^{d/2}}\sum_{n=0}^{\infty}\omega_{n}^{d-2}+\cr\cr
&-&\frac{c}{a}\frac{\Gamma\left(1-(d/2)\right)}{(4\pi)^{d/2}}\sum_{n=0}^{\infty}\omega_{n}^{d-2}+a\delta\Lambda\ .
\end{eqnarray}

	The surface counter-term $c$ is calculated considering the self-energy $\Sigma$ expressed in (\ref{Sigmac}), as well as the condition (\ref{condicaocorrecaoG}). In this case, these expressions are given, respectively, by
\begin{equation}
\label{SigmaDN}
\Sigma_{DN}(\xi)=4\lambda G_{DN}(\xi,\xi)+\delta m^{2}-c\delta(\xi_{d}-a)\ ,
\end{equation}
\begin{equation}
\label{rty5}
\int_{0}^{a}d\xi_{d}G_{DN}(x,\xi)\Sigma_{DN}(\xi)G_{DN}(y,\xi)\rightarrow finite\ .
\end{equation}

	As in the case of NN conditions, the integral (\ref{rty5}) is divergent due to the integration over the regions near the planes, $\xi=0,a$, so condition (\ref{rty5}) can be substituted by the equivalent one
\begin{equation}
\label{rty6}
\int_{a-\epsilon}^{a}d\xi_{d}G_{DN}(x,\xi)\Sigma_{DN}(\xi,\xi)G_{DN}(y,\xi)+\int_{0}^{\epsilon}d\xi_{d}G_{DN}(x,\xi)\Sigma_{DN}(\xi,\xi)G_{DN}(y,\xi)\rightarrow finite\ ,
\end{equation}
where $\epsilon$ is an infinitesimal positive arbitrary parameter.

	The second term on the left hand side of (\ref{rty6}) is finite. This can be shown by considering that it is a good approximation to take $\xi_{d}\sim0$ in this term, once the integration runs over $0\leq \xi_{d}\leq\epsilon$ and $\epsilon$ is infinitesimal. But, when $\xi_{d}\sim0$, it can be shown that $G_{DN}(x,\xi)\sim \xi_{d}$ and also $\Sigma_{DN}(\xi)\sim(\xi_{d})^{1-d}$. Therefore the second term on the left hand side of (\ref{rty6}) reads
\begin{eqnarray}
\label{rty7}
\int_{0}^{\epsilon}d\xi_{d}G_{DN}(x,\xi)\Sigma_{DN}(\xi)G_{DN}(y,\xi)&\sim&\int_{0}^{\epsilon}d\xi_{d}\ \xi_{d}(\xi_{d})^{1-d}\xi_{d}\sim\int_{0}^{\epsilon}d\xi_{d}\ (\xi_{d})^{3-d}\cr\cr
&\sim<&\int_{0}^{\epsilon}d\xi_{d}=finite\ ,
\end{eqnarray}
where, in the second line, we used that $d=2,4$. Equation (\ref{rty7}) states that the second term on the left hand side of (\ref{rty6}) is finite, and therefore, condition (\ref{rty6}) reads
\begin{equation}
\label{rty8}
\int_{a-\epsilon}^{a}d\xi_{d}G_{DN}(x,\xi)\Sigma_{DN}(\xi,\xi)G_{DN}(y,\xi)\rightarrow finite.
\end{equation}

	Using the fact that
\begin{equation}
\frac{\partial G_{DN}(x,\xi)}{\partial \xi_{d}}\Bigg|_{\xi_{d}=a}=0\ .
\end{equation}
we see that $G_{DN}(x,\xi)$ does not depend on $\xi_{d}$ near the plane $\xi_{d}=a$. Therefore, for a small $\epsilon$, condition (\ref{rty8}) reads
\begin{equation}
G_{DN}(x,\xi)|_{\xi_{d}=a}\left(\int_{a-\epsilon}^{a}d\xi_{d}\Sigma_{DN}(\xi,\xi)\right)G_{DN}(y,\xi)|_{\xi_{d}=a}\rightarrow finite\ .
\end{equation}
Taking into account the fact that $G_{DN}(x,\xi)|_{\xi_{d}=a}$ is finite, we have
\begin{equation}
\label{rty9}
\int_{a-\epsilon}^{a}d\xi_{d}\Sigma_{DN}(\xi,\xi)\rightarrow finite\ .
\end{equation}
Adding the finite contribution $\int_{a/2}^{a-\epsilon}d\xi_{d}\Sigma_{DN}(\xi,\xi)$ and introducing a convenient well behaved function (the cosine), the condition (\ref{rty9}) reads
\begin{equation}
\label{rty10}
\int_{a/2}^{a}dz^{d}\Sigma_{DN}(z,z)\cos\left(\frac{\pi z^{d}}{a}\right)\rightarrow finite\ .
\end{equation}

	The introduction of the cosine in (\ref{rty10}) is justified by the fact that, along the integration domain, it is a finite function with constant (negative) sign, and has the property of not depending on $\xi_{d}$ for $\xi_{d}\sim a$. Hence, it does not disturb the divergences which appear in (\ref{rty9}).

	Substituting (\ref{SigmaDN}) in (\ref{rty10}) and using equations (\ref{espcontgerl1}), (\ref{varphiDN}) and (\ref{qn;omegaDN}) we have
\begin{equation}
\label{poi1}
-4\lambda\frac{\Gamma\Bigl(1-(d/2)\Bigr)}{(4\pi)^{d/2}}\left[\frac{1}{4}\omega_{0}^{d-2}+\frac{1}{\pi}\sum_{n=0}^{\infty}\omega_{n}^{d-2}\right]-\delta m^{2}\frac{a}{\pi}-\frac{c}{2}\rightarrow\ finite\ ,
\end{equation}
where we used that $\int_{a/2}^{a}d\xi_{d}\delta(\xi_{d}-a)=-1/2$. Taking into account the definition of $\omega_{n}$ expressed in (\ref{qn;omegaDN}), the analytic extension (\ref{extanalitica4}) and the result (\ref{deltam2}), condition (\ref{poi1}) reads
\begin{eqnarray}
\label{poi2}
\left[-\lambda\frac{\Gamma\Bigl(1-(d/2)\Bigr)}{(4\pi)^{d/2}}\left(m^{2}+\left(\frac{\pi}{2a}\right)^{2}\right)^{(d-2)/2}+\frac{c}{2}\right]&+&\cr\cr
+\frac{-16\lambda}{(4\pi)^{(d+1)/2}}\frac{am^{d-1}}{\pi}\left[2\sum_{n=1}^{\infty}\frac{K_{(d-1)/2}(4man)}{(2man)^{(d-1)/2}}-\sum_{n=1}^{\infty}\frac{K_{(d-1)/2}(2man)}{(man)^{(d-1)/2}}\right]&\rightarrow&\ finite\ .
\end{eqnarray}

	Analogously to the NN case, the $\Gamma$ function in the first term on the left hand side of (\ref{poi2}) produces a divergent $a$-independent contribution for $d=2$, which is canceled out by taking
\begin{equation}
\label{defcDN}
c=2\lambda\frac{\Gamma\Bigl(1-(d/2)\Bigr)}{(4\pi)^{d/2}}m^{d-2}-\lambda m^{d-2}\sigma_{DN}(d)\ ,
\end{equation}
where $\sigma_{DN}(d)$ is an arbitrary dimensionless function of $d$, the product $m^{d-2}\sigma_{DN}(d)$ is finite and we took into account that $c$ can not depend on the distance $a$.

	For $d=3$ the $\Gamma$ function in (\ref{poi2}) is finite and the introduction of the surface counter-term $c$ is not necessary. Then we must take
\begin{equation}
\label{sigmad=3DN}
\sigma(d=3)=2\frac{\Gamma\Bigl(1-(3/2)\Bigr)}{(4\pi)^{3/2}}=-\frac{1}{2\pi}\ ,
\end{equation}
what makes $c$ equal to zero in (\ref{defcDN}) for $d=3$.

	Similarly to the NN case, for $d=2$ the function $\sigma_{DN}(d)$ can not be determined.

	The $\delta\Lambda$ counter-term present in (\ref{poi3}) is given by (\ref{dL}) and it is determined by the same arguments used in the previous sections.
	
	From equations (\ref{deltam2}), (\ref{dL}), (\ref{qn;omegaDN}), (\ref{poi3}) and (\ref{defcDN}), and also, using the analytic extension (\ref{extanalitica4}), we obtain
\begin{eqnarray}
\label{E1DNd}
{\cal E}^{1}_{DN}&=&\frac{2\lambda}{a}\Biggl[\frac{4am^{d-1}}{(4\pi)^{(d+1)/2}}\Biggl(2\sum_{n=1}^{\infty}\frac{K_{(d-1)/2}(4man)}{(2man)^{(d-1)/2}}-\sum_{n=0}^{\infty}\frac{K_{(d-1)/2}(2man)}{(man)^{(d-1)/2}}\Biggr)+\cr\cr
&+&\left(\frac{\Gamma(1-(d/2))}{4(4\pi)^{d/2}}m^{d-2}\left(\frac{\Gamma(1-(d/2))}{\Gamma(2-d)}-2\right)+\frac{m^{d-2}\sigma_{DN}(d)}{4}\right)\Biggr]^{2}+\cr\cr
&-&\frac{2\lambda}{a}\left[\left(\frac{\Gamma(1-(d/2))}{4(4\pi)^{d/2}}m^{d-2}\left(\frac{\Gamma(1-(d/2))}{\Gamma(2-d)}-2\right)+\frac{m^{d-2}\sigma_{DN}(d)}{4}\right)^{2}\right]\ ,
\end{eqnarray}
which is the first radiative correction to the Casimir energy in $d$ dimensions for DN conditions.

	For $d=2,3$ dimensions we have, respectively,
\begin{eqnarray}
\label{E1DNd=3}
{\cal E}^{1}_{DN}\Bigl|_{d=3}&=&\frac{\lambda m^{2}}{32\pi^{2}a}\left[\left(\frac{2}{\pi}\sum_{n=1}^{\infty}\frac{K_{1}(4man)-K_{1}(2man)}{n}\right)^{2}-1\right]
\ ,\\
\label{E1DNd=2}
{\cal E}^{1}_{DN}\Bigl|_{d=2}&=&\frac{\lambda\gamma^{2}}{32\pi^{2}a}\Biggl[\left[\frac{2}{\gamma}\ln\left(\frac{1-\exp(-2ma)}{1-\exp(-2ma)}\right)+\left(1+\frac{2\pi\sigma_{DN}(2)}{\gamma}\right)\right]^{2}+\cr\cr
&\ &-\left(1+\frac{2\pi\sigma_{DN}(2)}{\gamma}\right)^{2}\Biggl]\ ,
\end{eqnarray}
where we used that $\Gamma(-1/2)=-2\sqrt\pi$ and $\Gamma(-1)\to\infty$ as well as equation (\ref{sigmad=3DN}) for the result (\ref{E1DNd=3}). For the result (\ref{E1DNd=2}) we used equations (\ref{rty}) and (\ref{qweap1}).

	The zero mass limitof equation (\ref{E1DNd}) is given by
\begin{equation}
\label{E1DNm=0}
{\cal E}_{DN}\Big|_{m=0}=2\lambda a^{3-2d}\left[\frac{4}{(4\pi)^{(d+1)}}\left(2^{2-d}-1\right)\Gamma^{2}\left(\frac{d-1}{2}\right)\zeta^{2}(d-1)\right]\ .
\end{equation}
where we used equation (\ref{zxcap}) and the fact that $\lim_{m\rightarrow\infty}m^{d-2}=0$.

	For $d=2,3$, equation (\ref{E1DNm=0}) yields
\begin{eqnarray}
\label{E1DNm=0d=3}
{\cal E}_{DN}\Big|_{d=3,m=0}&=&\frac{\lambda}{2^{9}3^{2}a^{3}}\ ,\\
\label{E1DNm=0d=2}
{\cal E}_{DN}\Big|_{d=3,m=0}&=&\frac{\lambda}{2^{3}\pi^{2}a}\ln^{2}(2)\ ,
\end{eqnarray}
where for the result (\ref{E1DNm=0d=3}) we used that $\Gamma(1)=1$ and $\zeta(2)=\pi^{2}/6$, and for the result (\ref{E1DNm=0d=2}) we used that
\begin{equation}
\lim_{d\rightarrow2}\left[\frac{4}{(4\pi)^{(d+1)}}\left(2^{2-d}-1\right)\Gamma^{2}\left(\frac{d-1}{2}\right)\zeta^{2}(d-1)\right]=\left(\frac{\ln(2)}{4\pi}\right)^{2}\ .
\end{equation}

	Results (\ref{E1DNm=0d=3}) and (\ref{E1DNm=0d=2}) agree with those found in the literature \cite{NPnossos}.

	Similarly to what happened to the previous cases, result (\ref{E1DNm=0d=3}) could be obtained by taking the limit $m\to0$ in (\ref{E1DNd=3}) with the aid of (\ref{ert2}). The same procedure does not work for $d=2$ once this limit when applied to equation (\ref{E1DNd=2}) diverges. For $d=2$ we must take first $m\to0$ and then $d=2$ as we did to obtain (\ref{E1DNm=0d=2}).

\section{Conclusions and Final Remarks}

	In this paper, using dimensional regularization, we calculated the first radiative correction to the Casimir energy of a complex massive scalar field with a self-interaction term $\lambda\phi^{4}$ in $2+1$ and $3+1$ dimensions. The restriction of these specific dimensions is justifyed by the fact that for higher spatial dimensions the $\lambda\phi^{4}$ theory is not renormalizable, even without boundary conditions.
	
	We considered the field satisfying boundary conditions at two parallel planes. Specificaly, we studied four tipes of boundary conditions for the field, namely: quasi-periodic conditions (which interpolate continuously the periodic and anti-periodic cases), Dirichlet conditions, Neumann conditions and mixed conditions. In the zero-mass limit we obtained the results previously published in the literature \cite{Symanzik,KrechDietrichPRA92,NPnossos}. It is interesting to note that for a massless field the results for DD and NN conditions are equal for both cases of $3+1$ and $2+1$ dimensions.
	
	This was not expected since any two-loop contribution to the Casimir energy depends on the form of the field modes and not only on the corresponding eigenfrequencies as it occurs with the one-loop contribution to the Casimir effect, which can be obtained through the zero-point energy of the field.
	
	However, for the case of a massive field, the two-loop contribution to the Casimir effect are different for DD and NN conditions, as it can be seen from equations (\ref{E1DDd=3}), (\ref{E1DDd=2}), (\ref{E1NNd=3}) and (\ref{E1NNd=2}).  
	
	For configurations involving Neumann conditions in $2+1$ dimensions, one needs surface counter-terms, and the first radiative correction to the Casimir energy can not be completely determined. There remains an undetermined arbitrary factor introduced by the surface counter-terms. Maybe this fact is an indication of the limitation of idealized boundary conditions (in this case, Neumann condition) commonly used in the literature, or even a limitation of the meaning of $2+1$ models.
	
	The appearance of the undetermined function of the dimension $\sigma_{NN}(d)$ alerts us for other kind of problems that may appear when computing two-loop diagrams under Neumann boundary conditions. Maybe the regularization/renormalization program can not be achieved successfully with such an idealized boundary condition. This problem deserves further investigation and we think it is not a peculiarity of self-interacting scalar fields. It seams that first radiative corrections to the QED Casimir effect with infinitely permeable plates instead of perfectly conducting ones has unremovable divergences\footnote{This problem is under investigation at the moment, but preliminary results suggest that the two-loop Casimir energy of QED with permeable plates is ill defined \cite{Thiago}}.

\section*{Acknowledgments}

	F.A. Barone and C. Farina would like to thank FAPESP and CNPQ for financial support

\appendix
\section*{Appendix}

	In this appendix we calculate summations needed to obtain some results presented in the paper.

	It is well established in the literature that the analytic continuation of the so called Epstein function is given by \cite{ambjornwolframAP83}
\begin{eqnarray}
\label{extanalitica2}
\sum_{n=0}^{\infty}\Biggl[m^{2}+\Biggl(\frac{n\pi}{a}\Biggr)^{2}\Biggr]^{-s/2}&=&\frac{1}{2}m^{-s}+\frac{1}{2}\sum_{n=-\infty}^{\infty}\Biggl[m^{2}+\Biggl(\frac{n\pi}{a}\Biggr)^{2}\Biggr]^{-s/2}\cr\cr
&=&\frac{1}{2}m^{-s}+\frac{am^{1-s}}{2\sqrt{\pi}\Gamma(s/2)}\Biggl[\Gamma\Biggl(\frac{s-1}{2}\Biggr)+4\sum_{n=1}^{\infty}\frac{K_{(1-s)/2}(2man)}{(man)^{(1-s)/2}}\Biggr]\ ,
\end{eqnarray}
which is valid for any $s$ complex, except $s=1,-1,-3,-5,...$, where it has simple poles.

	From the previous equation, we also have the following analytical extension
\begin{eqnarray}
\label{extanalitica4}
\sum_{n=0}^{\infty}\Biggl[m^{2}+\Biggl(n+\frac{1}{2}\Biggr)^{2}\Biggl(\frac{\pi}{a}\Biggr)^{2}\Biggr]^{-s/2}&=&\sum_{n=0}^{\infty}\Biggl[m^{2}+\Biggl(\frac{n\pi}{2a}\Biggr)^{2}\Biggr]^{-s/2}-\sum_{n=0}^{\infty}\Biggl[m^{2}+\Biggl(\frac{n\pi}{a}\Biggr)^{2}\Biggr]^{-s/2}\cr\cr
&=&\frac{am^{1-s}}{2\sqrt{\pi}\Gamma(s/2)}\Gamma\Biggl(\frac{s-1}{2}\Biggr)+\cr\cr
&+&\frac{2am^{1-s}}{\sqrt{\pi}\Gamma(s/2)}\Biggl[2\sum_{n=1}^{\infty}\frac{K_{(1-s)/2}(4man)}{(2man)^{(1-s)/2}}+\cr\cr
&\ &-\sum_{n=1}^{\infty}\frac{K_{(1-s)/2}(2man)}{(man)^{(1-s)/2}}\Biggr]\ .
\end{eqnarray}

	Another important summation used in the paper is calculated using the fact that
\begin{equation}
K_{1/2}(2man)=\frac{1}{2}\frac{\sqrt{2\pi}}{\sqrt{2man}}e^{-2man}\ ,
\end{equation}
which allows us to write
\begin{eqnarray}
\label{somaparatheta}
&\ &\sum_{n=1}^{\infty}\cos(n\theta)\frac{K_{1/2}(2man)}{(man)^{1/2}}= \frac{\sqrt{\pi}}{4ma}\sum_{n=1}^{\infty}\Biggl[\frac{e^{-n(2ma-i\theta)}}{n}+\frac{e^{-n(2ma+i\theta)}}{n}\Biggr]\cr\cr
&=&\frac{\sqrt{\pi}}{4ma}\sum_{n=1}^{\infty}\Biggl[\int_{1}^{\infty}d\alpha(2ma-i\theta)e^{-\alpha n(2ma-i\theta)}+
\int_{1}^{\infty}d\alpha(2ma+i\theta)e^{-\alpha n(2ma+i\theta)}\Biggr]\cr\cr
&=&\frac{\sqrt{\pi}}{4ma}\Biggl[(2ma-i\theta)\int_{1}^{\infty}d\alpha \sum_{n=1}^{\infty}e^{-\alpha n(2ma-i\theta)}+
(2ma+i\theta)\int_{1}^{\infty}d\alpha \sum_{n=1}^{\infty}e^{-\alpha n(2ma+i\theta)}\Biggr]\cr\cr
&=&\frac{\sqrt{\pi}}{4ma}\Biggl[(2ma-i\theta)\int_{1}^{\infty}d\alpha\frac{e^{-\alpha(2ma-i\theta)}}{1-e^{-\alpha(2ma-i\theta)}}+
(2ma+i\theta)\int_{1}^{\infty}d\alpha\frac{e^{-\alpha(2ma+i\theta)}}{1-e^{-\alpha(2ma+i\theta)}}\Biggr]\cr\cr
&=&-\frac{\sqrt{\pi}}{4ma}\ln\Bigl(1+e^{-4ma}-2e^{-2ma}\cos(\theta)\Bigr)\ .
\end{eqnarray}
In the special case where $\theta=0$ we have
\begin{eqnarray}
\label{qweap1}
\sum_{n=1}^{\infty}\frac{K_{1/2}(2man)}{(man)^{1/2}}=-\frac{\sqrt{\pi}}{2ma}\ln\Bigl(1-e^{-2ma}\Bigr)\ .
\end{eqnarray}

	Along the text, in the zero mass limits, we used the result \cite{Arfken}
\begin{equation}
\lim_{z\rightarrow 0}K_{\nu}(z)\rightarrow\frac{2^{\nu-1}\Gamma(\nu)}{z^{\nu}}\ \ ,\ \ (\nu>0)
\end{equation}
in order to write
\begin{eqnarray}
\label{zxcap}
\lim_{m\rightarrow 0}\Biggl[m^{2\nu}\sum_{n=1}^{\infty}\frac{K_{\nu}(2man)}{(man)^{\nu}}\Biggr]&=&\sum_{n=1}^{\infty}\Gamma(\nu)m^{2\nu}\frac{2^{\nu-1}}{(2man)^{\nu}}\frac{1}{(man)^{\nu}}\cr\cr
&=&\frac{1}{2}\Gamma(\nu)\zeta(2\nu)\frac{1}{a^{2\nu}}\ ,
\end{eqnarray}
where $\zeta$ is the Riemann zeta-function. Analogously, we have
\begin{equation}
\label{zxcap1}
\lim_{m\rightarrow 0}\Biggl[m^{\nu}\sum_{n=1}^{\infty}\cos(n\theta)\frac{K_{\nu}(man)}{n^{\nu}}\Biggr]=2^{\nu-1}\Gamma(\nu)\frac{1}{a^{\nu}}\sum_{n=1}^{\infty}\frac{\cos{n\theta}}{n^{2\nu}}\ .
\end{equation}

	In the zero mass limit we have
\begin{equation}
\label{ert2}
\sum_{n=1}^{\infty}\frac{K_{1}(2man)}{man}\cong\frac{1}{2(ma)^{2}}\sum_{n=1}^{\infty}\frac{1}{n^{2}}=\frac{1}{2(ma)^{2}}\zeta(2)=\frac{\pi^{2}}{12(ma)^{2}}\ ,
\end{equation}
where we used that $K_{1}(x)\cong(1/x)$, as $x\to 0$.



\end{document}